1# Deep Learning-Enabled Swallowing Monitoring and Postoperative Recovery Biosensing System

Chih-Ning Tsai, Pei-Wen Yang, Tzu-Yen Huang, Jung-Chih Chen, Hsin-Yi Tseng, Che-Wei Wu, Amrit Sarmah and Tzu-En Lin****Abstract —*** **This study introduces an innovative 3D-printed dry electrode tailored for biosensing in postoperative recovery scenarios. Fabricated through a drop-coating process, the electrode incorporates a novel 2D material, MXene, and PEDOT:PSS on a polylactide (PLA) substrate. The PEDOT:PSS layer functions as an effective oxidation barrier for MXene, thereby enhancing the electrode's conductivity, biocompatibility, stability, and reusability. The design of the electrode is inspired by the paraboloidal dome-shaped suction cups found on tentacles of the octopus, a feature that substantially increases the surface area. These electrodes have been successfully integrated into a surface electromyography (sEMG) system, designed to monitor postoperative conditions in patients diagnosed with neck cancer or dysphagia. The system leverages a deep learning model to aid physicians in the quantitative assessment of post-surgical conditions of patients. Additionally, the study outlines a novel manufacturing approach for biosensing systems, demonstrating considerable promise in improving the utility in clinical environments.**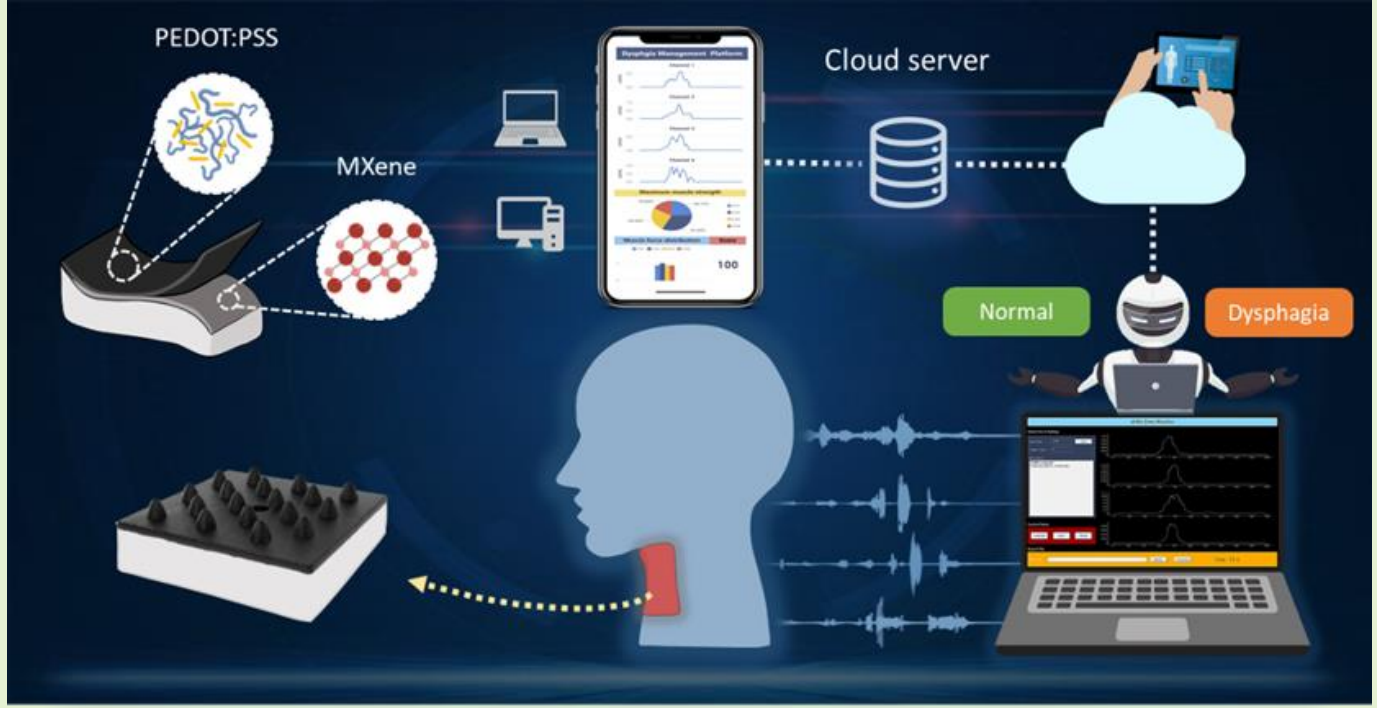

***Index Terms—*** **Dry Electrode, MXene, Surface Electromyography, Dysphagia, Swallowing Monitoring, Postoperative Recovery, Wearable Device, PEDOT:PSS**

## I. INTRODUCTION

DYSPHAGIA is a complex swallowing disorder that poses considerable diagnostic challenges in clinical settings. The disorder is associated with adverse outcomes, including malnutrition, dehydration, respiratory infections, prolonged hospital stays, and higher mortality rates from associated conditions [1]. The standard swallowing mechanism involves the action of the suprahyoid and infrahyoid muscle groups. These muscles facilitate the descent of the base and the elevation and forward movement of the larynx, ensuring smooth entry of the bolus into the esophagus. Food can remain in the pharynx or enter the trachea if this mechanism is disrupted, leading to aspiration [2]. Dysphagia can be caused by various factors, such as gastro-esophageal reflux disease and various cancers, leading to swallowing difficulties. For example, in the case of oropharyngeal dysphagia, oral transfer and pharyngeal phases of swallowing are affected.

Traditional diagnostic techniques primarily involve instrumental methods such as video fluoroscopic swallowing studies (VFSS) and fiberoptic endoscopic swallowing evaluations (FEES) [3]. However, these techniques have limitations. They often require specialized expertise and equipments so that they can be financially burdensome [4] [5] [6]. Given these constraints, comprehensive differential diagnosis methods, particularly in postoperative cases, have become essential for ongoing investigations [7]. To address the diagnostic challenges of dysphagia, this study focuses on the application of non-invasive surface electromyography (sEMG) techniques as innovative alternatives to traditional invasive electromyography [8] [9] [10] [11]. Unlike their invasive counterparts, sEMG electrodes facilitate a less intrusive evaluation of swallowing muscles, particularly their contractility and coordination, making them especially beneficial in preliminary evaluations of swallowing mechanisms [12] [13]. By integrating deep learning algorithms within a postoperative recovery biosensing system, we leverage the capabilities of sEMG to facilitate the early detection and assessment of dysphagia, thus offering valuable clinical insights. However, a recognized limitation of traditional sEMG is its susceptibility to signal interference attributed to the evaporation of the conductive gel on wet electrodes [14] [15] [16]. To mitigate this challenge, we introduce a novel dry electrode design in this work, aiming to improve signal fidelity and thus enhance the utility of sEMG in dysphagia diagnosis and evaluation of dysphagia.

Building upon existing literature in swallowing monitoring systems, which predominantly rely on traditional wet electrodes, strain sensors, or piezoelectric signals [4] [17], our study introduces a multichannel approach for enhanced dysphagia detection. While researchers have previously employed single-channel sEMG and machine-learning strain sensors to classify dysphagia [18], research focusing on using multichannel dry electrodes combined with deep-learning algorithms remains limited. We have developed a novel 3D-printed dry electrode capable of acquiring 4-channel sEMG signals to fill this gap. Utilizing octopus-inspired MXene nanosheets/PEDOT: PSS coated electrodes (OMP electrodes), our study aims to improve the accuracy of dysphagia detection by applying a deep learning system, specifically a Convolutional Neural Network (CNN). We implemented high-pass filters to preprocess the sEMG signals and remove noise to ensure real-time response. One of the salient features of our self-made dry OMP electrodes is their suitability for long-term wear, thereby overcoming the limitations of wet electrodes, which require an electrolyte gel that can solidify and degrade signal quality over time [19] [20] [21] [22] [23] [24] [25] [26] [27]. Furthermore, our dry electrodes enhance wearability and monitoring capabilities and offer cost-effectiveness and convenience compared to traditional wet electrode systems.

## II. EXPERIMENTS

### A. Fabrication of a 3D-printed OMP electrode

*1) Preparation of the OMP electrode:* The design of electrodes is essential to acquire accurate EMG recordings during signal acquisition. In this study, we draw inspiration from the unique suction cup found in the tentacles of an octopus, incorporating a paraboloidal dome-shaped artificial protrusion in the electrode design. This paraboloidal dome-like protrusion significantly contributes to the patch's increased surface area, enhancing the electrode's effectiveness in signal acquisition. We produced the OMP electrode using a commercial photocuring 3D printer (Phrozen Technology). We manufactured its substrates using PLA as the printing material. After completing the 3D printing process through photopolymerization, the electrode underwent ultrasonic cleaning for 5 minutes, followed by a secondary curing process. These steps were integral to developing the OMP electrode, ensuring its quality and suitability for our intended applications. Fig. 1(a) provides a detailed representation of the entire procedure.

The SEM image shows the material surface observed using a scanning electron microscope (SEM). This type of microscope employs an electron beam to scan the sample. It generates high-resolution images based on reflected electron signals, allowing for a detailed examination of the microstructure and surface morphology of the material. Typically, SEM images present highly magnified views that enable us to observe features at the micrometer or even nanometer scale. Fig. 1(b) illustrates the coating of MXene on the parabolic surface structure. Through the magnified observation of SEM images, we can examine the details of MXene nanosheets more closely. Fig. 1(b) also shows that the MXene nanosheets have a single-layer structure.

The preparation of the OMP electrode involved several steps. Initially, we subjected the electrode to ultrasonic cleaning in $H_2O$ and ethanol. To enhance the electrode's conductivity and biocompatibility, we applied MXene and PEDOT: PSS coatings to the surface of the PLA substrate. MXene ($Ti_3C_2T_x$), a group of two-dimensional metal carbides or nitrides, is used as a conductive connection layer to adhere PEDOT:PSS and an insulated polylactic acid (PLA) substrate. MXenes and PEDOT: PSS are highly conductive, biocompatible, and eco-friendly, making them ideal for wearable electronics and sEMG-based disease-detecting biosensors [28] [29] [30] [31] [32]. The process began by adding 25 μL of MXene solution (with a concentration of 8.52 mg/mL) onto the electrode surface and drying it in an oven at 50°C until fully dry. We drop-casted 10 μL of PEDOT: PSS onto the MXene/PLA surface, allowing it to dry and form the PEDOT: PSS/MXene/PLA layer. Finally, an additional 15 μL of PEDOT: PSS solution was added to the OMP electrode's surface and dried, as depicted in Fig. 1 (c). This comprehensive approach ensures the successful creation of the OMP electrode, optimizing its conductivity and biocompatibility for our intended applications.

*2) Surface and Contact Area Quantification:* We used Blender software to calculate the theoretical surface area of the OMP electrode as 122 mm$^2$, surpassing the 100 mm$^2$ of a flat electrode due to its artificial paraboloidal dome-shaped protrusion. Using ImageJ software after applying blue ink to the electrode and attaching it to the neck, we confirmed an actual contact area of 117.44 mm$^2$, thereby validating the protrusion's effectiveness in increasing surface area.

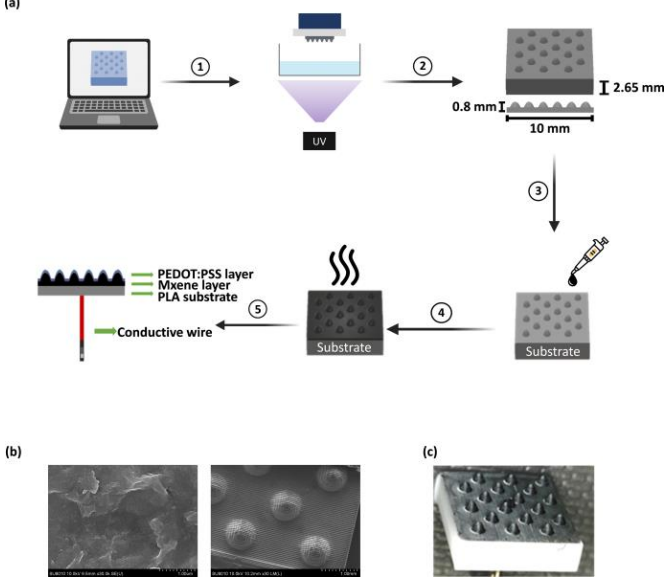

Fig. 1. Stages of Electrode Production: (a) Electrode substrate fabrication and coating process for electrode. (b) SEM images of MXene and layered structure schematic of OMP electrode (c) Surface photo featuring MXene and PEDOT:PSS coatings.

### B. Surface EMG Recording System

We have designed a surface electromyography (sEMG) recording system to efficiently capture muscle electrical signals and precisely measure the electrode's contact area with the skin. This system's design is crucial as it directly impacts the quality and reliability of the EMG signals. Regarding hardware design (Fig. 2(a)), we have employed the STM32 NUCLEO- L031K6 microcontroller, MyoWare muscle sensor, and HC-05 Bluetooth module. With a sampling rate of 250 Hz and a 12-bit resolution, we have also utilized DMA technology to enhance data transfer efficiency while reducing the microcontroller's load, ensuring efficient and uninterrupted data transfer from the STM32 to the computer during sEMG signal acquisition. Subsequently, we performed preprocessing on the acquired sEMG signals, including using an 8$^{th}$-order Butterworth high-pass filter to eliminate baseline noise, rectification, and root mean square (RMS) calculation with a 200-millisecond window. This preprocessing highlights the practical components of the EMG signals, providing reliable data for subsequent research and applications. Through this system and methodology, we accurately measure the contact area and capture high-quality sEMG signals, establishing a solid foundation for electromyography research and applications.

*1) Hardware design:* Fig. 2 (a) shows the sEMG recording system, which included an STM32 NUCLEO-L031K6 microcontroller, MyoWare muscle sensor, and HC-05 Bluetooth; data was wirelessly transmitted to a computer via Bluetooth [33] [34]. We set the sampling rate at 250 Hz and the resolution to 12 bits. We used the direct memory access (DMA) technique in hardware design to optimize the data transfer rate and reduce the microcontroller load [35] [36]. This approach ensured efficient and reliable data transfer from the STM32 to the computer without interrupting the sEMG signal acquisition process. As a result, we captured high-quality EMG signals with minimal noise and interference. Fig. 2 (b) illustrates the programming flow chart, which starts with the initialization of Direct Memory Access (DMA), Analog-to-Digital Converter (ADC), and Universal Asynchronous Receiver/Transmitter (UART). After this, we initiate two separate threads. The first thread collects the initial set of data from four ADCs and, upon completing data acquisition, raises the Complete Flag to a high state. The DMA controller then stores the ADC data into memory.

Conversely, the second thread accesses the following ADC data from memory after each transfer, continuing the data transmission sequentially. In the timing diagram depicted in Fig. 2 (c), UART initiates data transmission as soon as it retrieves ADC data from memory. This transmission continues until new ADC data becomes available, at which point the process repeats.

*2) Preprocessing:* As shown in Fig. 3 (a)-(b), we processed the obtained sEMG signals using an 8$^{th}$-order Butterworth high-pass filter with a 100 Hz cutoff frequency to eliminate baseline noise. After this step, we rectified the filtered signal and calculated the root mean square (RMS) using a 200-millisecond window size. [37] [38] [39].

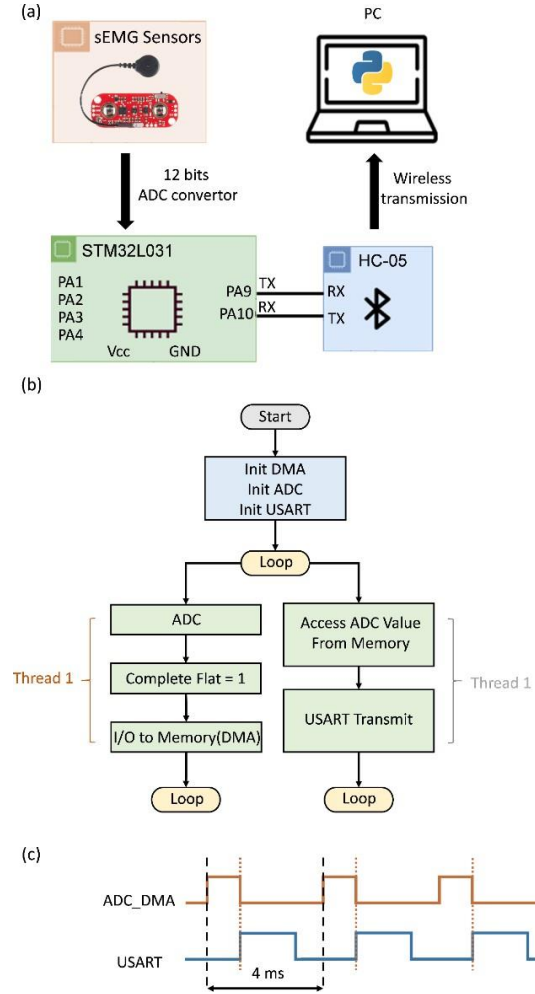

Fig. 2. Comprehensive overview of the semg recording system: schematic, programming flow, and timing. (a) Electrical schematic of sEMG recording system. STM32 Microcontroller Functions: (b) Programming flow chart and (c) Timing diagram. ("High" indicates a high voltage level, "Low" indicates a low voltage level.)

## III. RESULTS AND DISCUSSIONS

### A. Evaluation of OMP electrode coating materials and structures

To optimize the design of a dry electrode, the shape, and conductive coating materials are vital considerations. We investigated the efficacy of various coating materials applied to electrode surfaces and various structural designs for collecting sEMG signals. The materials we tested included carbon gel, commercial gel electrodes (Ag/AgCl), MXene, and PEDOT:PSS/MXene. Fig. 3 (c) shows the original signals obtained from various conductive materials, including Ag/AgCl, MXene, PEDOT:PSS/MXene, and carbon gel. Our findings indicate that among the tested materials, electrodes coated with MXene and PEDOT:PSS showed the best performance. These coatings exhibited lower noise in the time domain. They achieved the highest signal-to-noise ratio (SNR), as detailed in Fig. 3 (d).

Further, Fig. 3 (e) highlights the SNR metrics obtained using different electrode structures: the paraboloidal dome-shaped protrusion surface, flat surface, and traditional gel electrode. Regarding structural design, the paraboloidal dome-shaped protrusion surface offered a larger contact area than the flat surface, enhancing SNR performance [40] [41]. Our research demonstrates that the PEDOT:PSS/MXene coating material and the paraboloidal dome-shaped protrusion surface electrode structure outperform the evaluated materials and structures.

### B. Evaluation of Water Swallowing Signals and Sensitivity Tests of the sEMG system

To test the sEMG sensor system, we conducted water swallowing tests. Fig. 4 (a) illustrates that ingesting different water volumes—5 mL, 10 mL, and 15 mL—resulted in distinct sEMG signals due to throat muscle contractions. We employed commercial gel electrodes (represented in blue) and OMP electrodes (represented in orange) as sEMG sensors to capture action potentials in muscle fibers. Our results demonstrate a strong correlation between the amplitude of the RMS peak and the intensity of the sEMG signals, which varied according to the volume of water swallowed. Notably, the OMP electrodes displayed greater sensitivity in detecting the volume of water swallowed, as evidenced in Fig. 4(b).

### C. Stability Tests for Long-term Healthcare Monitoring

The OMP electrodes show considerable promise for long-term healthcare monitoring, capturing high-quality sEMG signals after 24 hours of uninterrupted use. As Fig. 4(c) demonstrates, the commercial wet Ag/AgCl electrode began to experience difficulties obtaining accurate signals just 4 hours into the monitoring period, and it completely lost its functionality by the next day. In stark contrast, the dry OMP electrodes maintained their performance throughout extended wear, as corroborated by this study.

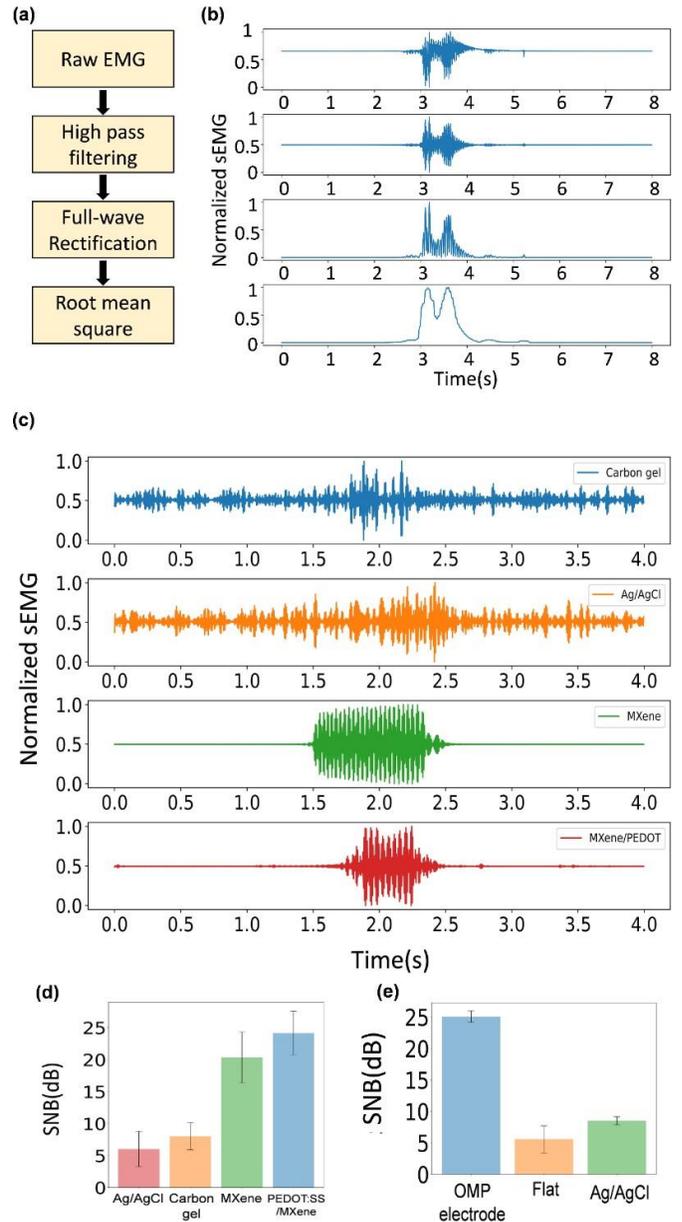

Fig. 3. Explanation of EMG signal preprocessing, including high-pass filtering, rectification, RMS computation, and notch filtering to eliminate powerline interference. (a) The signal preprocessing flow chart. (b) Waveforms in the time domain for each step. (c) Filtered sEMG signals were captured using a range of conductive coatings. (d) Comparative analysis of filtered sEMG signals and signal-to-noise ratio (SNR) across different coating materials and electrode structures. SNR evaluations correspond to various coating materials are displayed. (e) SNR performance across different electrode structures with PEDOT:PSS/Mxene and commercial gel electrodes.

### D. Skin Impedance Measurements

We evaluated electrical impedance spectroscopy (EIS) using an electrochemical analyzer, CHI627E (CH Instruments, Inc.). We conducted measurements in the frequency range from 10 to 1000 Hz. In Fig.4 (d), initially (0 hr), the OMP electrodes had impedances slightly higher than those of the commercial gel electrodes. However, after five hours, we observed a marked increase in impedance measured by the commercial gel electrodes compared to that of the OMP electrodes, as depicted in Fig. 4 (e), demonstrating the superior OMP electrode stability.

## E. Deep Learning-Enabled Postoperative Recovery Biosensing System for Dysphagia

To assess the performance of the postoperative biosensing system for dysphagia, we collected data from fourteen participants. Seven participants were healthy individuals (four men and three women) between the ages of 22 and 27 without a history of neurological disorders or throat surgeries. The other seven participants were patients with dysphagia (six men and one female) aged between 35 and 80 years old. All participants wore electrodes while swallowing motions to measure sEMG data in four channels. Fig. 5 (a)(b) shows the four-channel sEMG waveforms of healthy individuals and patients with dysphagia. In Fig. 5 (b), it is evident that patients find it challenging to perform a complete swallowing motion due to swallowing difficulties. Consequently, there is a recurring attempt to swallow, resulting in a longer duration for eating than that of individuals without such problems.

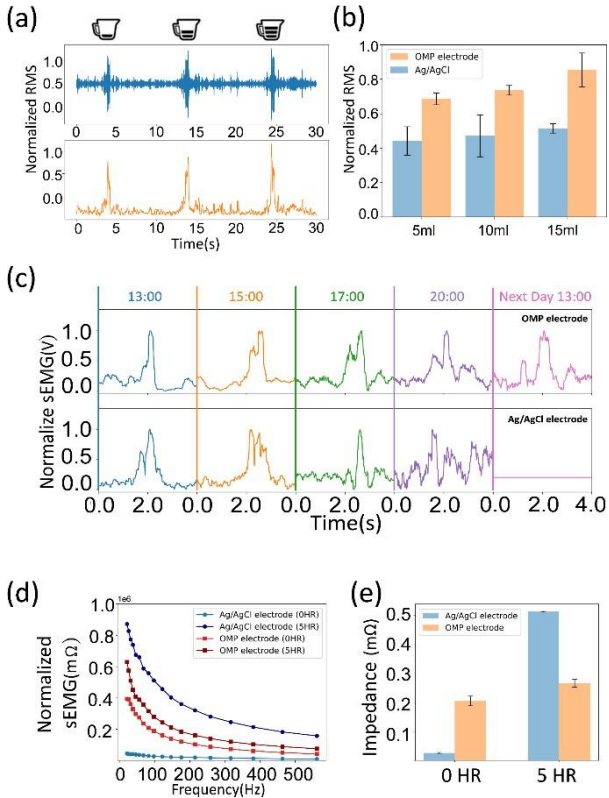

Fig. 4. Evaluation of water swallowing signals and sensitivity tests of the sEMG system commercial gel electrodes (Ag/AgCl) and the OMP electrodes. Stability assessment for long-term healthcare monitoring using sEMG sensors developed in this work. (a) The RMS values were obtained by swallowing 5 ml, 10 ml, and 15 ml, respectively. (b) Corresponding RMS values for commercial gel electrodes (Ag/AgCl) and the OMP electrodes. (c) A full day of sEMG monitoring using OMP electrodes. (d) Comparison of impedance magnitude versus frequency for both electrode types, measured at time zero and five hours later. (e) Fluctuations in skin-electrode contact impedance were assessed at 100 Hz using commercial Ag/AgCl and dry OMP electrodes.

Data from dysphagia patients was collected to train a CNN model for classifying sEMG signals. The model took a one-dimensional time series signal from a swallow motion and processed it through three convolutional and pooling layers sections to extract features. Convolutional layers used ReLU activation followed by maximum pooling, preventing overfit- ting. Extracted features were flattened and connected to a fully connected layer after batch normalization. The output layer employs a sigmoid function, constraining the output range between 0 and 1, making it interpretable as probabilities allow us to represent the probability of the dysphagia level. Suppose the probability value output by the sigmoid function is closer to 1. In that case, it implies a higher likelihood of resembling a patient and receiving a lower score. Conversely, suppose the value is closer to 0. In that case, it suggests a more remarkable resemblance to an average person, resulting in a higher score. We call this probability value the 'Health Index.' A higher Health Index indicated a healthier state, while a lower Health Index suggested a greater similarity to a state associated with dysphagia. If the probability value output by the sigmoid function was closer to 1, it implied a higher likelihood of resembling a patient and receiving a lower score. Conversely, if the value was closer to 0, it suggested a more remarkable resemblance to an average person, resulting in a higher score. Fig. 5 (c)-(d) illustrates the training and validation performance of the 1D-CNN model over 200 iterations. The loss rapidly decreases within the initial iterations, and further training leads to even lower loss, indicating successful convergence. The absence of overfitting is evident from consistent trends in accuracy and loss between training and validation sets, showcasing the model's generalization ability. The ROC curve (Fig.5(e)) displays an impressive AUC of 98.33%. The final classifier's confusion matrix (Fig. 5 (f)) demonstrates a 98.84% diagnostic accuracy, 98.24% precision, and 100% recall for two categories, reflecting robust and reliable classification performance.

To test various forms of input data, we employed a 2D-CNN that utilizes the short-time Fourier transform (STFT) to generate spectrograms containing both time and frequency domain information. Finally, we summarize the results of 1D-CNN and 2D-CNN in TABLE I, showing that 1D-CNN has higher accuracy, AUC, precision, sensitivity, and F1 score than 2D-CNN. For one-dimensional time series signals, 1D-CNN has better feature extraction capability than 2D-CNN because there is no signal processing method such as the short-time Fourier transform (STFT), or wavelet transform (WT) before CNN. Direct input of the raw time series signal ensures the authenticity of the input. Compared to 2D-CNN, 1D-CNN only performs one-dimensional convolution, making the structure more straightforward and requiring fewer parameters. As a result, it can save computational resources and time and is more conducive to real-time machine monitoring.

Finally, a cloud-based management platform designed to allow users and physicians to easily monitor the swallowing status, highlighting the potential of intelligent healthcare solutions. Fig. 5 (g) shows that the platform architecture involves data collection using electrodes, data transmission through Bluetooth to a computer for processing, and deep learning-assisted diagnosis is performed. The data and diagnostic results are synchronized and uploaded to a MySQL database for storage. The stored data can be accessed using various devices, such as smartphones, tablets, and computers, using the cloud platform for communication and retrieval. Fig. 5 (h) shows the results displayed on the screen in a graphical format, including waveform data, muscle force analysis, and muscle recovery status. This system supports long-term monitoring and treatment, allowing patients to assess their swallowing status at home and decreasing the need for frequent monitoring clinic visits. Physicians can access patient data anytime for efficient and informed decision-making. The intelligent system provides real-time monitoring for innovative healthcare care, positioning it as a promising tool for monitoring and managing swallowing disorders.

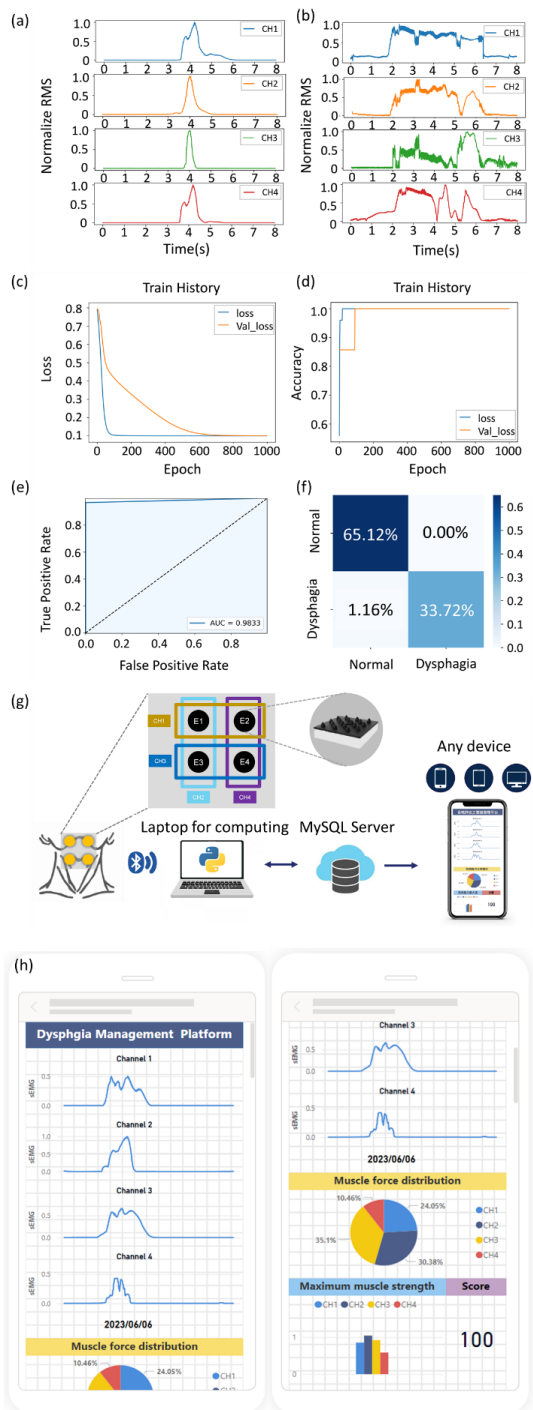

Fig. 5. (a) shows the four-channel sEMG waveforms of healthy individuals and (b) shows the four-channel sEMG waveforms of patients with dysphagia. (c) The loss function and (d) classification accuracy are plotted against the number of iterations. The curves demonstrate that the model has converged successfully without overfitting, indicating the effectiveness of the proposed model design. (e) Evaluation of the performance of the 1D-CNN model using external data. ROC curve with an AUC of 98.33%. (f) Confusion matrix of the best classifier. The overall diagnostic accuracy is 98.84%, with a precision of 98.24% and a recall of 100% for both categories. (g) A cloud-based platform for monitoring swallowing facilitates accessible data collection, transmission, and AI-assisted diagnosis, supporting long-term monitoring. This platform empowers patients and enables efficient decision-making for physicians, making it a promising tool for managing swallowing disorders. (h) The mobile interface allows users to conveniently access the system's features and recovery scores on their smartphones.

## IV. CONCLUSIONS

We have developed a novel deep learning-enabled swallowing monitoring and postoperative recovery biosensing system in this investigation. Our approach involved refining both the OMP electrode design and signal processing methodologies to ensure the acquisition of high-quality sEMG signals. We coated the paraboloidal dome-shaped electrode with MXene and PEDOT: PSS to enhance conductivity and biocompatibility. Moreover, we systematically improved signal processing strategies, employing advanced filtering and noise reduction techniques and fine-tuning the deep learning model. This comprehensive approach resulted in 98.84% precision, 100% sensitivity, and 98.24% precision in the evaluation of dysphagia, significantly outperforming previous studies in the field. The model can effectively distinguish swallowing patterns using only a 4-channel sEMG, eliminating the need for multiple sensors or noise-prone microphones that are commonly used in other AI-related swallowing research [17] [18] [19]. Ultimately, it shows promise for clinical implementation in assessing swallowing function and diagnosing dysphagia. This innovative method can enhance patient care and provide healthcare professionals with an efficient and reliable diagnostic tool.

Further studies must thoroughly explore its capabilities and validate its effectiveness in real-world clinical settings. Moreover, this system incorporates a cloud platform that allows a detailed graphical representation of swallowing information includes assessing any damage, comparing the effectiveness of different treatments, and allowing long-term continuous follow-up to improve the patient's understanding of their swallowing conditions. In future work, individual experiments will address the limited datasets available.

## ACKNOWLEDGMENT

This work was supported by the Young Scholar Fellowship Program of the NSTC in Taiwan [grant numbers 112-2636-E-A49 -007] and [112-2221-E-A49 -051 -].

## ETHICAL CONSENT

Institutional Review Board of Kaohsiung Medical University Hospital, Reference number KMUHIRB-E(I)-20230179 granted ethical approval for the involvement of human subjects in this study.